\documentclass{ifcolog}

\usepackage{amsmath}
\usepackage{amssymb}

\newbox\tempa
\newbox\tempb
\newdimen\tempc
\def\mud#1{\hfil $\displaystyle{\mathstrut #1}$\hfil}
\def\rig#1{\hfil $\displaystyle{#1}$}
\def\irulehelp#1#2#3{\setbox\tempa=\hbox{$\displaystyle{\mathstrut #2}$}%
                        \setbox\tempb=\vbox{\halign{##\cr
        \mud{#1}\cr
        \noalign{\vskip\the\lineskip}
        \noalign{\hrule height 0pt}
        \rig{\vbox to 0pt{\vss\hbox to 0pt{${\; #3}$\hss}\vss}}\cr
        \noalign{\hrule}
        \noalign{\vskip\the\lineskip}

        \mud{\copy\tempa}\cr}}
                      \tempc=\wd\tempb
                      \advance\tempc by \wd\tempa
                      \divide\tempc by 2 }
\def\irule#1#2#3{{\irulehelp{#1}{#2}{#3}
                     \hbox to \wd\tempa{\hss \box\tempb \hss}}}

\begin{document}
\title{Rules and derivations in an elementary logic course}
\titlerunning{Rules and derivations in an elementary logic course}
\addauthor{Gilles Dowek}{Inria, 23 avenue d'Italie, CS 81321, 
75214 Paris Cedex 13, France. {\tt gilles.dowek@inria.fr}}
\authorrunning{Gilles Dowek}
\titlethanks{}
\maketitle

When teaching an elementary logic course to students who have a
general scientific background but have never been exposed to logic, we
have to face the problem that the notions of deduction rule and of
derivation are completely new to them, and are related to nothing they
already know, unlike, for instance, the notion of model, that can be
seen as a generalization of the notion of algebraic structure.

In this note, we defend the idea that one strategy to introduce these
notions is to start with the notion of inductive definition
\cite{Aczel}.  Then, the notion of derivation comes naturally.  We
also defend the idea that derivations are pervasive in logic and that
defining precisely this notion at an early stage is a good investment
to later define other notions in proof theory, computability theory,
automata theory, ... Finally, we defend the idea that to define the
notion of derivation precisely, we need to distinguish two notions of
derivation: {\em labeled with elements} and {\em labeled with rule
names}.  This approach has been taken in \cite{Dowek}.

\section{From inductive definitions to derivations}

\subsection{A method to define sets: inductive definitions}

Inductive definitions are a way to define subsets of a set $A$. The
definition of a subset $P$ is given by functions $f_1$, from $A^{n_1}$
to $A$, $f_2$, from $A^{n_2}$ to $A$, ...  These functions are called
{\em rules}.  For example, the function $f_1 = \langle \rangle \mapsto
0$, from ${\mathbb N}^0$ to ${\mathbb N}$, and $f_2 = \langle a
\rangle \mapsto a + 2$, from ${\mathbb N}^1$ to ${\mathbb N}$ are
rules.

Instead of writing these rules $f_1 = \langle \rangle \mapsto 0$ and
$f_2 = \langle a \rangle \mapsto a + 2$, we often write them
$$\irule{}{0}{f_1}$$
$$\irule{a}{a+2}{f_2}$$
but despite this new notation, rules are things the students know:
functions.

To define the subset $P$, we first define a function $F$ from ${\cal
  P}(A)$ to ${\cal P}(A)$ as follows
$$F(X) = \bigcup_i \{f_i(a_1,..., a_{n_i})~|~a_1, ..., a_{n_i} \in X\}$$
For example, the two rules above define the function 
$$F(X) = \{0\} \cup \{a + 2~|~a \in X\}$$
and, for instance, $F(\{4, 5, 6\}) = \{0, 6, 7, 8\}$, $F(\varnothing)
= \{0\}$, $F(\{0\}) = \{0, 2\}$, ...

The function $F$ monotonic and continuous, thus it has a smallest
fixed point $P$ which is the inductively defined subset of $A$.  This
smallest fixed point can be defined in two ways
$$P = \bigcap_{F(X) \subseteq X} X = \bigcup_i F^i(\varnothing)$$ 
The first definition characterizes the set $P$ as the smallest set
closed by $f_1$, $f_2$, ... the second as the set containing all the
elements that can be built with these functions in a finite number of
steps.  The monotonicity and continuity of $F$ and the two fixed points
theorems are easy lemmas \cite{Dowek} for mathematically oriented
students.  They can be admitted without proof otherwise.

Continuing with our example, the set $P$ of even numbers can
be characterized as the smallest set containing $0$ and closed by the
function $a \mapsto a + 2$, or as the union of the sets
$F(\varnothing) = \{0\}$, $F^2(\varnothing) = \{0, 2\}$,
$F^3(\varnothing) = \{0, 2, 4\}$, ...

\subsection{Derivations}

Defining a {\em derivation} as a tree whose nodes are labeled with
elements of $A$ and such that if a node is labeled with $x$ and its
children with $y_1$, ..., $y_n$, then there exists a rule $f$ such
that $x = f(y_1, ..., y_n)$, and {\em a derivation of an element $a$}
as a derivation whose root is labeled with $a$, we can prove, by
induction on $i$, that all the elements of $F^i(\varnothing)$ have a
derivation.  The property is trivial for $i = 0$. If it holds for $i$
and $a \in F^{i+1}(\varnothing)$, then by definition $a = f(b_1, ...,
b_n)$ for some rule $f$ and $b_1 \in F^i(\varnothing)$, ..., $b_n \in
F^i(\varnothing)$, thus, by induction hypothesis, $b_1$, ..., $b_n$
have derivations. Hence, so does $a$.

Thus, from the second property $P = \cup_i F^i(\varnothing)$, we get
that all elements of $P$ have derivations. Conversely, all elements
that have a derivation are elements of $P$.

Continuing with our example the number $4$ has the derivation
$$\irule{\irule{\irule{} 
                      {0}
                      {} 
               }
               {2}
               {}
        } 
        {4}
        {}$$

\subsection{Rule names}

There are several alternatives for defining the notion of
derivation. For instance, when $x = f(y_1, ..., y_n)$, instead of
labelling the node just with $x$, we can label it with the ordered
pair formed with the element $x$ and the name of the rule $f$.  For
instance, the derivation of $4$ above would then be
$$\irule{\irule{\irule{}
                      {\langle 0, f_1\rangle}         
                      {}
               }
               {\langle 2, f_2\rangle}
               {}
        }
        {\langle 4, f_2\rangle}
        {}$$
more often written 
$$\irule{\irule{\irule{}
                      {0}         
                      {f_1}
               }
               {2}
               {f_2}
        }
        {4}
        {f_2}$$
Such a derivation is easier to check, as checking the node 
$$\irule{2}{4}{}$$ 
requires to find the rule $f$ such that $f(2) = 4$, while checking the
node
$$\irule{2}{4}{f_2}$$
just requires to apply the rule $f_2$ to $2$ and check that the result
is $4$.

But these rules names are redundant, as soon as the relation $\cup_i
f_i$ is decidable.  So, in general, they can be omitted.

\subsection{Derivations and derivations}

Instead of omitting the rule names, it is possible to omit the
elements of $A$.  The derivation of $4$ is then
$$\irule{\irule{\irule{}
                      {f_1}         
                      {}
               }
               {f_2}
               {}
        }
        {f_2}
        {}$$
that can also be written 
$$\irule{\irule{\irule{}
                      {~.~}
                      {f_1}         
               }
               {~.~}
               {f_2}
        }
        {~.~}
        {f_2}$$

Although it is not explicit in the derivation, the element $4$ can be
inferred from this derivation with a top-down {\em conclusion
inference algorithm}, because the rules $f_i$ are functions. The
conclusion of the rule $f_1$ can only be $f_1(\langle \rangle) = 0$,
that of the first rule $f_2$ can only be $f_2(\langle 0 \rangle) = 2$,
and that of the second can only be $f_2(\langle 2 \rangle) = 4$.

We can introduce this way two kinds of derivations {\em labeled with
objects} and {\em labeled with rules names}.

\subsection{Making the rules functional}

Natural deduction proofs, for instance, are often labeled both with
sequents and rule names
$$\irule{\irule{} 
               {P, Q, R \vdash P}
               {\mbox{axiom}}
         ~~~~~~~~~~~~~~~~~~~~~~~~~~~~~~~
         \irule{}
               {P, Q,R \vdash Q}
               {\mbox{axiom}}
        }
        {P, Q, R \vdash P \wedge Q}
        {\mbox{$\wedge$-intro}}$$
but they can be labeled with sequents only 
$$\irule{\irule{} 
               {P, Q, R \vdash P}
               {}
         ~~~~~~~~~~~~~~~~~~~~~~~~~~~~~~~
         \irule{}
               {P, Q,R \vdash Q}
               {}
        }
        {P, Q, R \vdash P \wedge Q}
        {}$$
and proof-checking is still decidable.  They can also be labeled with
rule names only, but we have to make sure that all the deduction rules
are functional, which is often not the case in the usual presentations
of Natural deduction.  The rule
$$\irule{\Gamma \vdash A~~~\Gamma \vdash B}
        {\Gamma \vdash A \wedge B}
        {\mbox{$\wedge$-intro }}$$
is functional: there is only one possible conclusion for each sequence
of premises, but the axiom rule
$$\irule{}{\Gamma, A \vdash A}{\mbox{axiom}}$$
is not. To make it functional, we must introduce a different rule
axiom$_{\langle \Gamma, A \rangle}$ for each pair $\langle \Gamma, A
\rangle$. Thus, the proof above must be written
$$\irule{\irule{} 
               {~.~}
               {\mbox{axiom$_{\langle \{Q, R\}, P \rangle}$}}
         ~~~~~~~~~~~~~~~~~~~~~~~~~~~~~~~
         \irule{}
               {~.~}
               {\mbox{axiom$_{\langle \{P, R\}, Q \rangle}$}}
        }
        {~.~}
        {\mbox{$\wedge$-intro}}$$
and its conclusion $P, Q, R \vdash P \wedge Q$ can be inferred
top-down.

\section{Derivations in elementary computability theory}

\subsection{A pedagogical problem}

The set of computable functions is often defined inductively as the
smallest set containing the projections, the null functions, and the
successor function, and closed by composition, definition by
induction, and minimization.

But to study the computability of properties of computable functions,
we need a notion of {\em program}, that is we need a way to express each
computable function by a expression, to which a G\"odel number can be
assigned. A usual solution is to introduce Turing machines at this
point.

This solution however is not pedagogically satisfying as, while the
students are still struggling to understand the inductive definition,
we introduce another, that is based on completely different ideas,
letting them think that logic made of odds and ends. Moreover, the
equivalence of the two definitions requires a tedious proof. Why do we
not base our notion of program on the inductive definition itself?

\subsection{Programs already exist}

The function $x \mapsto x + 2$ is computable because it is the
composition of the successor function with itself
$$\irule{\irule{}{x \mapsto x + 1}{}
         ~~~~~~~~~~~~~~~
         \irule{}{x \mapsto x + 1}{}
        }
        {x \mapsto x + 2}
        {}$$
But such a derivation labeled with objects cannot be used as a
program, because to label its nodes, we would need a a language to
express all the functions, and there is no such language.

But if we use a derivation labeled with rule names instead 
$$\irule{\irule{}{~.~}{Succ}
         ~~~~~~~~~~~~~~~
         \irule{}{~.~}{Succ}
        }
        {~.~}
        {\circ_1^1}$$
and write the trees in linear form: $\circ_1^1(Succ,Succ)$, we obtain
a simple functional programming language to express the programs.

For instance, introduce a G\"odel numbering $\ulcorner . \urcorner$
for these programs, and assume there is an always defined function $h$
such that
\begin{itemize}
\item $h(p, q) = 1$ if $p = \ulcorner f \urcorner$ and $f$ defined 
at $q$ 
\item $h(p, q) = 0$ otherwise 
\end{itemize}
then, the function
$$k = \circ_1^1 (\mu^1(\pi^2_1), \circ_2^1(h, \pi_1^1, \pi_1^1))$$
is defined at $\ulcorner k \urcorner$ if and only if it is not. 

We get this way a proof of the undecidability of the halting problem
that requires nothing else than the inductive definition of the set of
computable functions.

\section{Derivations in elementary automata theory}

When introducing the notion of automaton, we often introduce new
notions, such as those of transition rules and recognizability. Having
introduced the notion of derivation from the very beginning of the
course permits to avoid introducing these as new notions.

Consider for instance the automaton
$$odd \xrightarrow{~a~} even$$
$$even \xrightarrow{~a~} odd$$
where the state $even$ is final. In this automaton, the word $aaa$ is
recognized in $odd$.  Indeed
$$odd \xrightarrow{~a~} even \xrightarrow{~a~} odd \xrightarrow{~a~} even$$
If, instead of introducing a new notion of transition rule, we just
define transition rules as deduction rules
$$\begin{array}{ccc}
~~~~~
\irule{even}{odd}{a}
~~~~~
&
~~~~~
\irule{odd}{even}{a}
~~~~~
&
~~~~~
\irule{}{even}{\varepsilon}
~~~~~
\end{array}$$
then, the element $odd$ has a derivation
$$\irule{\irule{\irule{\irule{}
                             {even}
                             {\varepsilon}
                      }
                      {odd} 
                      {a}
               }
               {even}
               {a}
         }
         {odd}
         {a}$$
If we label this derivation with rule names we obtain
$$\irule{\irule{\irule{\irule{}
                             {~.~}
                             {\varepsilon}
                      }
                      {~.~} 
                      {a}
               }
               {~.~}
               {a}
         }
         {~.~}
         {a}$$
which can be written in linear form $a(a(a(\varepsilon)))$, or $aaa$.

Thus, a word $w$ is recognized in a state $s$ if and only if it is a
derivation, labeled with rule names, of $s$.

This example introduces some points to be discussed: the rules
$$\begin{array}{cc}
~~~~~
\irule{even}{odd}{a}
~~~~~
&
~~~~~
\irule{odd}{even}{a}
~~~~~
\end{array}$$
are labeled with the same name. If the automaton is deterministic, 
we can replace these two rules with one: a function such that 
$a(even) = odd$ and $a(odd) = even$. But for non deterministic 
automata, we either need to extend the notion of rule name, allowing 
different rules to have the same name, or to consider two rule names
$$\begin{array}{ccc}
~~~~~
\irule{even}{odd}{a_1}
~~~~~
&
~~~~~
\irule{odd}{even}{a_2}
~~~~~
&
~~~~~
\irule{}{even}{\varepsilon}
~~~~~
\end{array}$$
and map the derivation 
$a_1(a_2(a_1(\varepsilon)))$ to the word $a(a(a(\varepsilon)))$
with the function $|.|$ defined by: $|\varepsilon| = \varepsilon$, 
$|a_1(t)| = a(|t|)$, and $|a_2(t)| = a(|t|)$.

\section{Introducing the Brouwer-Heyting-Kolmogorov correspondence}

\subsection{A radical change in viewpoint?}

The Brouwer-Heyting-Kolmogorov interpretation, and its counterpart,
the Curry-de Buijn-Howard correspondence, are often presented as a
radical change in viewpoint: proofs are not seen as trees
anymore, but as algorithms.

But, of course, these algorithms must be expressed in some
language---often the lambda-calculus.  Thus, proofs are not really
algorithms, but terms expressing algorithms, and such terms are
nothing else than trees. So, it is fairer to say that, in the
Brouwer-Heyting-Kolmogorov interpretation, proofs are not derivation
trees, but trees of a different kind. For instance, the tree
$$\irule{\irule{\irule{\irule{}
                             {P \wedge Q \vdash P \wedge Q}
                             {}
                      }
                      {P \wedge Q \vdash Q} 
                      {}
                 ~~~~~~~~~~~~~~~~~~~~~~
                \irule{\irule{}
                             {P \wedge Q \vdash P \wedge Q}
                             {}
                      }
                      {P \wedge Q \vdash P } 
                      {}
               }
               {P \wedge Q \vdash Q \wedge P}
               {}
        }
        {\vdash (P \wedge Q) \Rightarrow (Q \wedge P)}
        {}$$
is replaced by the tree
$$\irule{\irule{\irule{\irule{}
                             {x}
                             {}
                      }
                      {fst} 
                      {}
                 ~~~~~~~~~~~~~~~~~~~~~~
                \irule{\irule{}
                             {x}
                             {}
                      }
                      {snd} 
                      {}
               }
               {\langle , \rangle}
               {}
        }
        {\lambda x: P \wedge Q}
        {}$$
often written in linear form: 
$\lambda x : P \wedge Q~\langle snd(x), fst(x) \rangle$.  

\subsection{What about derivation trees labeled with rule names?}

Instead of following this idea of expressing proofs as
algorithms, let us just try to label the derivation above with rule
names. Five rules are used in this proof. Three of them are
functional 
$$\irule{\Gamma \vdash A~~~\Gamma\vdash B}
        {\Gamma \vdash A \wedge B} 
        {\mbox{$\wedge$-intro}}$$
$$\irule{\Gamma \vdash A \wedge B}
        {\Gamma \vdash A}
        {\mbox{$\wedge$-elim1}}$$
$$\irule{\Gamma \vdash A \wedge B}
        {\Gamma \vdash B}
        {\mbox{$\wedge$-elim2}}$$
Let us just give them shorter names: $\langle , \rangle$, 
$fst$, and $snd$. The rule 
$$\irule{\Gamma, A \vdash B}
        {\Gamma \vdash A \Rightarrow B}
        {\mbox{$\Rightarrow$-intro}}$$
is functional, as soon as we know which proposition $A$ in the 
left-hand side of the antecedent is used. 
So, we need to supply this proposition $A$ in the rule name, 
let us call this rule $\lambda A$. Finally, the rule 
$$\irule{}
        {\Gamma, A \vdash A}
        {\mbox{axiom}}$$
is functional, as soon as we know $\Gamma$ and $A$. 
We could supply $\Gamma$ and $A$ in the rule name. However, we shall 
just supply the proposition $A$ and infer the context $\Gamma$. 
Let us call this rule $[A]$. 
So, the proof above can be written
$$\irule{\irule{\irule{\irule{}
                             {P \wedge Q \vdash P \wedge Q}
                             {[P \wedge Q]}
                      }
                      {P \wedge Q \vdash Q} 
                      {snd}
                 ~~~~~~~~~~~~~~~~~~~~~~
                \irule{\irule{}
                             {P \wedge Q \vdash P \wedge Q}
                             {[P \wedge Q]}
                      }
                      {P \wedge Q \vdash P } 
                      {fst}
               }
               {P \wedge Q \vdash Q \wedge P}
               {\langle , \rangle}
        }
        {\vdash (P \wedge Q) \Rightarrow (Q \wedge P)}
        {\lambda P \wedge Q}$$
and if we keep rule names only 
$$\irule{\irule{\irule{\irule{}
                             {~.~}
                             {[P \wedge Q]}
                      }
                      {~.~}
                      {snd}
                 ~~~~~~~~~~~~~~~~~~~~~~
                \irule{\irule{}
                             {~.~}
                             {[P \wedge Q]}
                      }
                      {~.~}
                      {fst}
               }
               {~.~}
               {\langle , \rangle}
        }
        {~.~}
        {\lambda P \wedge Q}$$
or in linear form $\lambda P \wedge Q~\langle snd([P \wedge Q]), 
fst([P \wedge Q]) \rangle$. This is the scheme representation 
\cite{DJ} of this proof.  

Let us show that the conclusion can be inferred, although we have not
supplied the context $\Gamma$ in the axiom rule. The conclusion
inference goes in two steps. First we infer the context 
bottom-up, using the fact that the conclusion
has an empty context, and that all rules preserve the context, except
$\lambda A$ that extends it with the proposition $A$
$$\irule{\irule{\irule{\irule{}
                             {P \wedge Q \vdash .}
                             {[P \wedge Q]}
                      }
                      {P \wedge Q \vdash .}
                      {snd}
                 ~~~~~~~~~~~~~~~~~~~~~~
                \irule{\irule{}
                             {P \wedge Q \vdash .}
                             {[P \wedge Q]}
                      }
                      {P \wedge Q \vdash .}
                      {fst}
               }
               {P \wedge Q \vdash .}
               {\langle , \rangle}
        }
        {\vdash .}
        {\lambda P \wedge Q}$$
Then, the right-hand part of the sequent can be inferred with a 
usual top-down inference algorithm, using the fact that the rules are
functional 
$$\irule{\irule{\irule{\irule{}
                             {P \wedge Q \vdash P \wedge Q}
                             {[P \wedge Q]}
                      }
                      {P \wedge Q \vdash Q}
                      {snd}
                 ~~~~~~~~~~~~~~~~~~~~~~
                \irule{\irule{}
                             {P \wedge Q \vdash P \wedge Q}
                             {[P \wedge Q]}
                      }
                      {P \wedge Q \vdash P}
                      {fst}
               }
               {P \wedge Q \vdash Q \wedge P}
               {\langle , \rangle}
        }
        {\vdash (P \wedge Q) \Rightarrow (Q \wedge P)}
        {\lambda P \wedge Q}$$

\subsection{Brouwer-Heyting-Kolmogorov interpretation: an optional modification}

In the rule 
$$\irule{\Gamma, A \vdash B}
        {\Gamma \vdash A \Rightarrow B}
        {\mbox{$\Rightarrow$-intro}}$$
instead of supplying just the proposition $A$, we can supply the 
proposition $A$ and a name $x$ for it. Then, in the axiom rule 
$$\irule{}{\Gamma, A \vdash A}{\mbox{axiom}}$$ 
instead of supplying the proposition $A$, we can just supply the name 
that has been introduced lower in the tree for it. We obtain this way 
the tree 
$$\irule{\irule{\irule{\irule{}
                             {~.~}
                             {x}
                      }
                      {~.~}
                      {snd}
                 ~~~~~~~~~~~~~~~~~~~~~~
                \irule{\irule{}
                             {~.~}
                             {x}
                      }
                      {~.~}
                      {fst}
               }
               {~.~}
               {\langle , \rangle}
        }
        {~.~}
        {\lambda x:P \wedge Q}$$
in linear form $\lambda x : P \wedge Q~\langle snd(x), fst(x)
\rangle$, which is exactly the representation of the proof according
to the Brouwer-Heyting-Kolmogorov interpretation.

So, the Brouwer-Heyting-Kolmogorov interpretation boils down to use of
derivations labeled with rule names plus two minor modifications:
context inference and the use of variables. These two modifications
can be explained by the fact that Natural deduction does not really
deal with sequents and contexts: rather with propositions, but,
following an idea initiated in \cite{SchoederHeister}, 
some rules such as the introduction rule of the implication dynamically add
new rules, named with variables.

\end{document}